# Defects, band bending and ionization rings in MoS$_2$

Iolanda Di Bernardo[1,2], James Blyth[1,2], Liam Watson,[1,2] Kaijian Xing,[2] Yi-Hsun Chen,[1,2] Shao-Yu Chen[1,2], Mark T. Edmonds[1,2,3] and Michael S. Fuhrer[1,2,3]

[1] Australian Research Council Centre of Excellence in Future Low-Energy Electronics Technologies, Monash University, Clayton, 3800, VIC, Australia
[2] School of Physics and Astronomy, Monash University, Clayton, 3800, VIC, Australia
[3] Monash Centre for Atomically Thin Materials, Monash University, Clayton, 3800, VIC, Australia

E-mail: iolanda.dibernardo@monash.edu, michael.fuhrer@monash.edu

**Abstract**

Chalcogen vacancies in transition metal dichalcogenides are widely acknowledged as both donor dopants and as a source of disorder. The electronic structure of sulphur vacancies in MoS$_2$ however is still controversial, with discrepancies in the literature pertaining to the origin of the in-gap features observed via scanning tunneling spectroscopy (STS) on single sulphur vacancies. Here we use a combination of scanning tunnelling microscopy (STM) and STS to study embedded sulphur vacancies in bulk MoS$_2$ crystals. We observe spectroscopic features dispersing in real space and in energy, which we interpret as tip position- and bias-dependent ionization of the sulphur vacancy donor due to tip induced band bending (TIBB). The observations indicate that care must be taken in interpreting defect spectra as reflecting in-gap density of states, and may explain discrepancies in the literature.

## 1. Introduction

Transition metal dichalcogenides, such as molybdenum disulfide (MoS$_2$), are attracting increasing attention as two-dimensional device components. Their layered structure allows for the easy isolation of atomically thin flakes which are suitable for a broad range of applications, including in electronics and optoelectronics.[1–3]

Crucial to the performance of devices based on the electronic properties of TMDs are defects in the lattice structure, with chalcogen vacancies being the most common ones.[4] They give rise to in-gap states[1,5–7] and affect the transport,[1] optical,[8,9] and catalytic properties of the device.[10,11] Sulfur vacancies in MoS$_2$, in particular, are responsible for the ubiquitous n-doping observed in the experiments. The role of S vacancies has been extensively studied in MoS$_2$ monolayers,[4,12–17] with some efforts being even directly devoted to the intentional introduction of defects by ion irradiation and thermal treatments.[4,18]

Density functional theory calculations consistently predict S vacancies in MoS$_2$ to yield an in-gap state, appearing as extra density of states (DOS) ~0.6 eV below the conduction band maximum (CBM). This feature corresponds to a doubly degenerate unoccupied state.[13–15,17,19] Additionally, occupied states are predicted to exist close to the valence band minimum (VBM).[15,20,21] Experimental measurements of the local density of states with scanning tunnelling spectroscopy (STS), on the other hand, have reported defect-induced resonances as DOS features located at a range of energies between the CBM and the VBM.[19,22,23]

However, STS on semiconductors can be complicated by several issues. Tip-induced band bending[2,24–27] (TIBB) occurs at a metal/vacuum/semiconductor junction, intrinsic to the configuration of STS measurements on semiconductors. The band bending locally perturbs the electronic structure of the sample and can give rise to errors in the estimation of bandgap size and defect binding energy.

Furthermore, TIBB may move defect states across the Fermi level, changing the defect occupancy, i.e. ionizing it. This tip position- and bias-dependent effect can lead to spurious features in STS described as ionization rings.[25,28–30] Related phenomena have been observed for other semiconductors and Dirac materials,[28,30,31] and very recently in line defects or grain boundaries in $MoS_2$ itself, where they give rise to quantization phenomena.[32]

In this work, we investigate native defects on the surface of cleaved, synthetic $MoS_2$, via scanning tunnelling microscopy and spectroscopy. We observe spectroscopic features within the bulk bandgap in the proximity of the defects whose energy varies spatially, suggesting that TIBB is indeed important in $MoS_2$. We attribute the observed features to the ionization of donors induced by TIBB and provide a simple electrostatic model to support our interpretation. Our results point out that caution must be exercised in interpreting spectroscopy features induced by defects in $MoS_2$.

## 2. Methods

Synthetic $MoS_2$ crystals were obtained by HQ graphene and kept in a nitrogen-filled glove box before usage. Before all measurements, the crystal was cleaved to expose a fresh surface.

Scanning tunnelling spectroscopy measurements were collected with a PtIr tip on the $MoS_2$ crystal cleaved in ultra-high vacuum, at a pressure better than $10^{-9}$ mbar. During measurements the sample temperature was kept at 77 K (liquid nitrogen). Spectroscopy data were collected by recording the tunnelling current while sweeping the bias, to record current vs. bias curves, then taking their derivative. STM and STS data were processed with the WSxM software.[33]

X-ray photoelectron spectroscopy (XPS) was performed on a Thermo Scientific Nexsa Surface Analysis System equipped with a hemispherical analyser. The incident radiation was monochromatic Al Kα X-rays (1486.6 eV) at 72 W (6 mA and 12 kV, 400 × 800 μm² spot). Survey (wide) and high-resolution (narrow) scans were recorded at analyser pass energies of 150 and 50 eV and step sizes of 1.0 eV and 0.1 eV, respectively. The base pressure in the analysis chamber was less than $5.0 \times 10^{-9}$ mbar. Data were analysed with the XPST tool for Igor Pro, using symmetric Voight lineshapes for the core level fitting.[34]

The AFM images are acquired by a Park XE-100 AFM with tapping mode, with a resolution of 256 sample/line.

Confocal Raman spectroscopy is performed in ambient with a homemade microscope. The sample is excited by a diode-pump solid-state continuous-wave laser at 532 nm (Coherent Sapphire SF 532-100). The excitation power is below 1 mW to avoid heating and sample degradation. A 100X objective lens (numerical aperture: 0.8) is employed to focus the laser down to the diffraction limit and collects the back-scattered light. The Raman signal is first filtered out by an edge filter (Semrock LP03-532RE), dispersed by a monochromator with an 1800 gr/mm grating (Princeton Instruments SP-2750), and then detected by a thermal-electrical-cooled charge-coupled device (Princeton Instruments PIXIS 256).

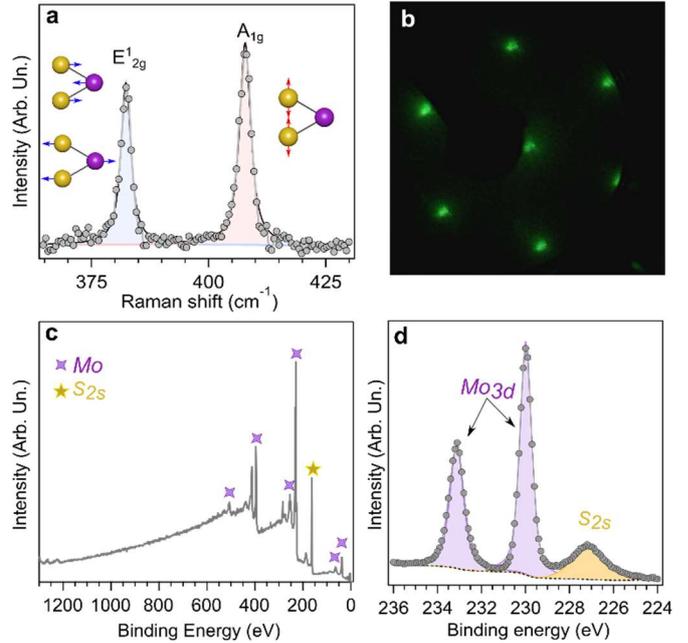

**Figure 1:** Large-area characterization of the $MoS_2$ surface. (a) Raman spectrum. (b) LEED pattern, beam energy 97eV. (c) XPS survey scan for elemental composition. (d) High-resolution XPS spectra of the Mo3d and S2s core levels.

For transport measurements $MoS_2$ flakes were exfoliated by the conventional scotch tape exfoliation technique onto a polydimethylsiloxane stamp. The exfoliated $MoS_2$ flake was then laminated onto the $SiO_2$/Si substrate by the dry transfer technique. After defining the electrode windows by direct-write photolithography (Intelligent micropatterning SF100 XPRESS), 60 nm Au was deposited by an electron-beam evaporator. The standard lift-off was processed by soaking the sample in PG remover at 60 ºC for 30 mins. Figure S2a depicts the $MoS_2$ Hall bar device. A Hall effect system (Janis) integrated with a coil magnet was employed to characterize the transport properties of the $MoS_2$ Hall bar device at room temperature. In the Janis system, a constant DC current of 100 nA was supplied by a Keithley source meter 2450 to the Hall-bar device with a perpendicular sweeping magnetic field from -0.7 T to + 0.7 T. Then the Hall voltage, $V_{Hall}$, was measured by an Agilent 34410A multimeter. The sample thickness was measured via AFM, and the carrier density was calculated as $n = I_{ds}/(etV_{Hall}/B_\perp)$, where $I_{ds}$ is the source-drain current, $e$ is the elementary charge, $t$ is the thickness of the flake, $V_{Hall}$ is the Hall voltage and $B_\perp$ is the magnetic field.

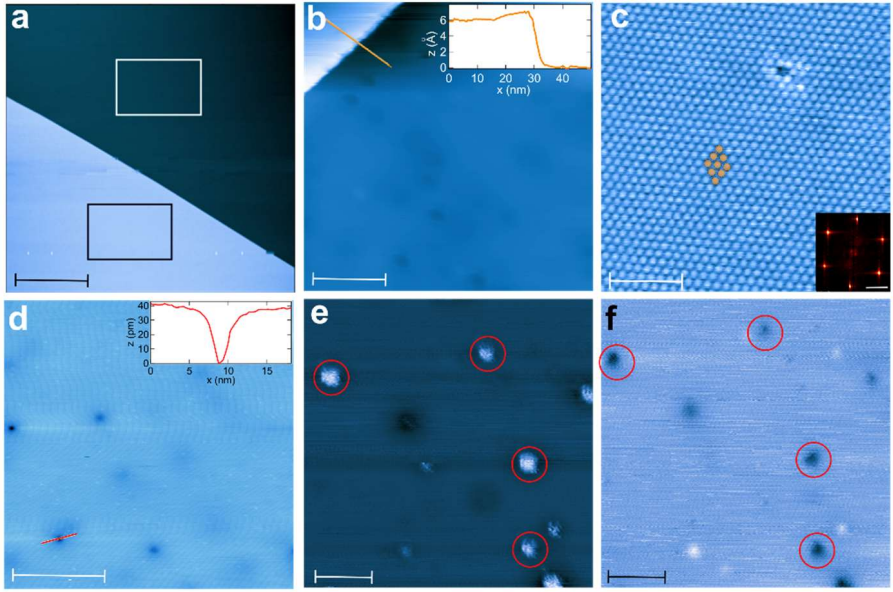

**Figure 2:** Scanning probe characterization of the cleaved 2H-MoS$_2$ (0001) surface. (a) (20 x 20) μm$^2$ AFM scan of freshly cleaved MoS$_2$, scalebar = 5μm. RMS roughness in the white box = 0.312 nm; RMS roughness in the black box = 0.307 nm. (b) (200 x 200) nm$^2$ STM scan of UHV-cleaved MoS$_2$, scalebar = 50 nm; I = 50 pA; V = -2 V. Inset: line profile across the orange line, corresponding to a step edge. Step height = 6.5 Å. (c) Atomic resolution STM scan showing a single atomic vacancy, scalebar = 2.5 nm; I = 50 pA; V = -1.8 V. Inset: FFT of panel (c) scalebar = 15 nm$^{-1}$. (d) (150 x 150) nm$^2$ scan, scalebar = 50 nm; I = 50pA; V = +1.5V. Inset: line profile across the red line in panel (d). (e) and (f): (50 x 50) nm$^2$ scans of the same area with opposite bias: (e) - 1.5V; (f) 1.5 V. For both scans, I = 50 pA; scalebar = 10 nm.

## 3. Results and discussion

### 3.1 Large scale surface characterization

Figure 1 reports the large-scale characterization of the MoS$_2$ sample. Figure 1a shows the Raman spectrum of the crystal excited with a 532 nm line in ambient conditions. Two main components can be identified in the spectrum, located at 383 and 408 cm$^{-1}$ respectively. These features correspond to the E$^1_{2g}$ (in-plane vibration of S atoms with respect to Mo) and A$_{1g}$ (out-of-plane oscillation of S atoms in opposite directions) non-resonant vibrational modes.[35–37] The 25 cm$^{-1}$ separation between the peaks is in line with the expected values for bulk MoS$_2$ crystals.[35,37]

The diffraction pattern acquired on the sample via low-energy electron diffraction (LEED) shows a single set of hexagonal spots, indicating high crystallinity over a large area (LEED spot size ~ 0.5 cm).

X-ray photoemission spectroscopy (XPS) measurements were carried out to verify the stoichiometry of the sample. In the survey spectrum, reported in Fig. 1c, only Mo and S components can be identified, besides common contaminants (C, O) attributed to air exposure. Fig. 1d shows the high-resolution spectrum of the Mo3d spin-orbit split doublet and the adjacent S2s core level. The centroids of the synthetic components for Mo3d$_{5/2}$ and S2p are found at binding energies of 230.0 and 227.1 eV respectively, in good alignment with previous reports.[38–41] The binding energy splitting between Mo3d$_{5/2}$ and S2p$_{3/2}$ core levels (extracted from Fig. 1c), sensitive to the exact MoS$_2$ stoichiometry, is found to be 67.2 eV, pointing out to a stoichiometry close to the nominal 1:2 ratio.[38]

### 3.2 Defects analysis

The surface of the freshly cleaved MoS$_2$ crystals was investigated with a combination of scanning probe techniques, to assess its homogeneity and estimate the concentration of intrinsic defects. Fig. 2a shows a (20$x$20) μm$^2$ AFM scan across a step edge. At this scale, the surface appears largely flat, with a root mean square (RMS) roughness of about 0.31 nm.

The same crystal was subsequently re-cleaved in UHV for scanning tunnelling microscopy (STM) analysis. Fig.1b is a representative large-scale scan of a freshly exposed surface. The most predominant feature in this panel is a step-edge visible in the top-left corner. A line profile across the step (orange line, reported in the inset of Fig. 1b) reveals a height of 0.65 nm, consistent with the interlayer spacing of a single S-Mo-S unit.[18,37] On both sides of the step the sample appears rather homogeneous, with some randomly distributed, slightly darker areas visible on the bottom terrace. The (10 x 10) nm$^2$ close-up on a flat area in Fig. 1c (I = 50 pA; V = -1.8V) reveals the atomically resolved hexagonal lattice of the sample. The lattice constant is measured to be $a = 3.15$ Å (superimposed orange model), in line with previous report on non-strained 2H-MoS$_2$ probed via STM.[18,19,42] The inset in Fig.1c shows the fast Fourier transform of the atomically resolved scan. We expect at this high bias that the features are purely topographic, i.e. only the top layer of sulfur atoms can be probed,[18] therefore the defect visible in panel 2c is attributed to a single sulfur vacancy on the topmost chalcogen layer. This interpretation is corroborated by the brighter appearance of the surrounding S atoms, caused by a charge redistribution around the unsaturated Mo dangling bonds.[13,14,16,42] Due to its low activation energy, this is indeed the most common type of defect observed on the MoS$_2$ surface.[4,13]

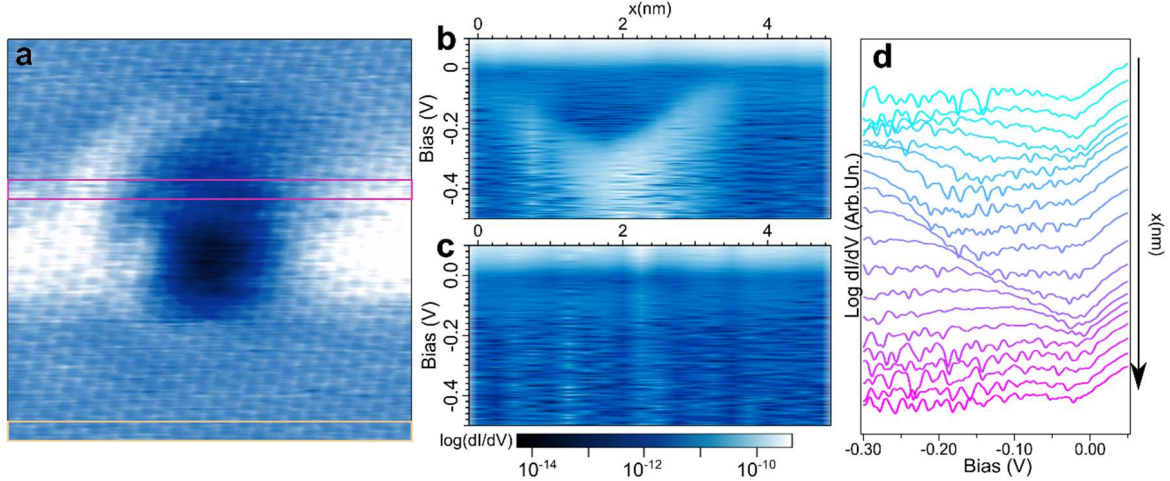

**Figure 3:** Spectroscopic characterization of the defect. (a) (**5 x 5**) nm$^2$ topographic image of the defect; I = 100 pA; V = 0.3 V. (b) and (c) STS spectra as a function of position, acquired along lines marked in pink and yellow, respectively, in panel (a). (c) Log-scale stack of the STS spectra for the line across the defect. Spectra are vertically offset for clarity.

Single sulfur vacancies are not, however, the only type of defects observed on the surface. As seen in Fig. 1d, randomly distributed dark areas, larger than single vacancies, can be spotted across the sample. A cross-sectional analysis (inset) over a typical "depression" reveals a depth of about 40 pm, a centrosymmetric appearance, and a diameter of about 4 nm. Given their larger radius, spanning several lattice constants, these larger features have previously been ascribed to sub-surface/buried defects.[22] By averaging on several large-area STM scans, covering a total area of 77500 nm$^2$, the concentration of surface defects is estimated to be $N_{2D} = 1.1 \times 10^{11} cm^{-2}$. Assuming a homogeneous distribution of such intrinsic defects across layers, and using a value for the interplanar distance $c = 6.5$ Å, we estimate $N_{3D} = N_{2D}/c = 1.7 \times 10^{18} cm^{-3}$. This value is in reasonable agreement with the n-type carrier density measured via Hall effect on the crystal (see Fig.S2), yielding $N = 1.3 \times 10^{19} cm^{-3}$.

Buried vacancies exhibit the same depression-like appearance regardless of the sign of the bias. We also observe, in some areas of the sample, defects that appear brighter or darker when imaged at positive or negative bias respectively. Examples of such defects are highlighted with red circles in Figures 2e and 2f. These defects, that change apparent height/depth as a function of the applied bias, are similar to those observed on Ar$^+$ irradiated MoS$_2$ crystals.[4,18,19] In previous work this type of bias-dependent behaviour has been attributed to sulfur multi-vacancies[19] or areas with a higher Mo concentration in general like Mo clusters,[43] as their radius extends well beyond the single unit cell. It is worth noting that the diameter of these features does not vary as a function of the scanning bias, rather showing a simple "switching" behaviour.

Intriguingly, some buried defects like the one reported in Fig. 3a appear surrounded by a circular, brighter "halo", surrounding a disc of lower signal intensity spanning a couple of nanometers. An atomically resolved image of the area rules out the presence of surface vacancies or other topographic surface features at the center of the disks. To shed light on the nature of these circular features and their dependence on bias and spatial coordinates, we collected a spatially resolved differential conductance map by acquiring STS spectra in a grid made of steps of 0.25 nm.

Spatially-resolved constant bias cuts in the range $[-0.03; -0.280]$ eV, well within the MoS$_2$ bandgap where no density of states features are expected, show "rings" of increased differential conductance. The diameter of the rings increases as a function of the applied voltage (Figure S2). The rings disappear at ~ -30 meV; above this energy the differential conductance maps appear featureless as expected. Given the bias dependence of this density of states feature, we infer its electronic rather than topographic origin.

To follow the spatial dependence of these features two exemplary line profiles of the conductance maps, taken and in the middle of (pink box) and away-from (yellow box in Fig. 3a) the disk, are shown in Fig. 3b and 3c, respectively. The horizontal and veritcal axes correspond to horizontal spatial coordinate and bias respectively. The colorscale, from dark to bright, reflects the logarithm of the differential conductance intensity. A vertical stack of the log-scale line profiles used for Fig. 3b is reported in Figure 3d.

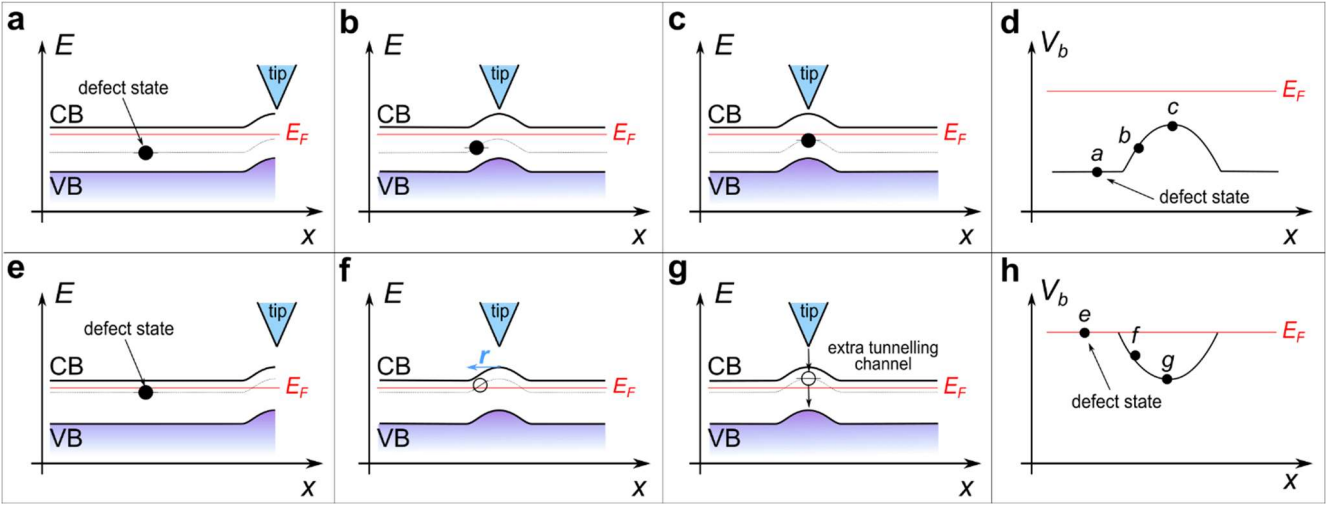

**Figure 4:** Schematic of the effect of TIBB in real space for a localised defect and in the case of ionization ring. Top row: cartoon model of the band configuration for a localized defect in the case of the tip being far (a), in proximity of (b) and directly above (c) the defect. The corresponding dispersion is pictured in (d). Bottom row: cartoon model of the formation of ionization rings as the tip is far away (e), in proximity of (f) and directly above (g) the defect. In (g) the donor level is pulled above the Fermi level $E_F$, and an extra tunnelling channel opens. The corresponding dispersion is pictured in (h).

The high intensity area visible at the top of the differential conductivity maps is due to the onset of the conduction band. From the location of the conduction band edge, just above the Fermi level $E_f$, we infer an overall n-type doping,[44] coherently with the observation of sulfur vacancies as the main type of defects and the Hall effect measurements (Fig. S1).

In panel 3b we trace the spatial evolution of the in-gap states (left to right or right to left). The in-gap density of states observed would suggest the presence of a defect. However, this spectral feature is found at different binding energies as a function of the distance from the centre of the defect. The cross section at a given energy (corresponding to a horizontal cut in panel 3b) is a ring-shaped feature, which explains the ring seen in Fig. 3a. The energy position of the feature is deepest (240 meV) when the diameter of the ring is smallest, and it gradually moves upwards towards the Fermi level, with the ring expanding, to then disappear about 1.8 nm away on either side. Coherently with this observation, the line profile corresponding to the edge of the scan, located 2.5nm away from the centre of the disc, appears completely featureless in this energy region.

### 3.3 Interpretation and Modelling

The presence of an impurity such as a dopant or a vacancy is known to create additional states in the bandgap of the materials, which can appear as in-gap states in the LDOS. For the following interpretation of the "dispersion" of defect states, we assume the observed defect to be a buried sulfur vacancy (i.e., a donor), as these are the most common type of defects in $MoS_2$.

We first consider the possibility that the observed behaviour is simply due to a "regular" donor state, and that the observed dispersion in energy and space is a consequence of TIBB alone. This scenario is schematically described in the top row of Figure 4. At zero bias between tip and sample, the work function of the tip (gold-coated PtIr, 5.1 eV[45]) is larger than the electron affinity of the n-type $MoS_2$ (-4.3 eV[46]). Bringing them within tunnelling distance from each other results in the formation of a junction, causing the formation of a charge depletion region in the $MoS_2$. This translates in the $MoS_2$ bands bending up in energy near the tip. An in-gap occupied defect state would be unperturbed when the tip is far away (Fig.4a) from it. With the tip approaching the defect laterally the donor level would be progressively pulled up in energy together with the bands because of TIBB (Fig. 4b). The donor level would be at its shallowest position in energy for the tip sitting exactly on the defect (Fig. 4c). In this situation, we would observe the donor level moving in energy as a function of the tip position, but this dipersion trend would appear as an arc with the concavity pointing downwards (Fig. 4d).

Our data displays the exact opposite trend. We attribute the observed behaviour to a tip position- and bias-induced ionization of the defect state instead. Similar bias- and spatial-dependent features in d$I$/d$V$ to Fig. 3 associated with the switching of a dopant (or vacancy) charge state caused by tip-induced band-bending (TIBB) were previously observed for Co adatoms on graphene,[30] Fe-doped Bi2Se,[25] Mn-doped[28][47] or "grained"[29] InAs and GaAs[48,49]. Interestingly, here we observe these patterns to be generated by a vacancy rather than a substitutional atom.

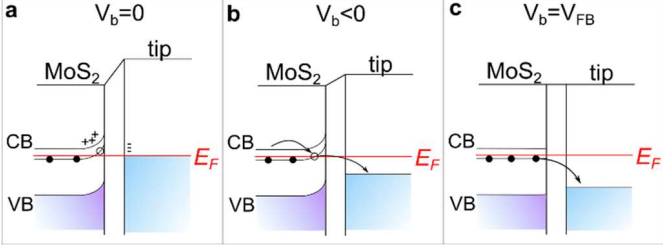

**Figure 5:** Schematic of the evolution of the band diagram at the tip-sample junction when the STM tip located on top of the MoS2 defect, as a function of the tip-sample bias. (a) Band bending at zero bias and formation of a depletion zone. (b) Conditions for the formation of the extra tunnelling channel, corresponding to the onset of the ionization ring. (c) Flat band condition.

Figures 4e-g show a schematic model of the lateral profile of the band bending condition as a function of the bias between tip and sample in this scenario. At zero bias, the lateral movement of the tip still induces a local band banding, pulling the MoS2 bands up in energy (Fig. 4e). In this case, at some distance $r$ from the tip the donor level is pulled above $E_f$, becoming unoccupied (ionized; Fig. 4f). Defects lying within the circle of radius $r$ from the tip will be positvely charged (emptied of an electron), while those laying outside will retain their filled condition. The screening from other electrons, in fact, masks off this perturbation at sufficiently large distances. When the defect is located within this ionization ring an additional tunnelling channel opens, corresponding to electrons tunnelling onto the (now empty) donor level first and then onto the empty states in the tip (Fig. 4g). This sudden, step-like change in the tunnelling current $I(V)$ is naturally accompanied by a peak-like feature in the $dI/dV$. The negative bias needed to see the onset of the extra current is largest when the tip is directly above the defect, as the donor level is pushed up in energy the most at that tip position. Overall, this produces a dispersion in space with the concavity pointing onwards (Fig. 4h).

Within this framework, we can also explain the observed dispersive trend as a function of the bias (i.e., changing ring diameter), as shown schematically in Fig. 5.

1. At zero (or positive tip bias) the Mos2 bands and the donor level are pulled up in energy, as shown in Fig. 5a. A depletion region is created at the MoS2 surface under the tip. The only conductivity is provided by bulk states and electrons tunnel from the filled states in the tip to the empty states of the sample.

2. When a negative bias is applied to the tip (Fig. 5b) the electric field between the tip and the sample is reduced, and the tip valence band is pushed below the Fermi level. Electrons from the MoS2 conduction band can now tunnel into the empty defect state and from there into the tip valence band, so effectively a new tunnelling channel is created. The exact bias required to push the defect below the Fermi level depends on the tip-defect distance, resulting in the observed ring-like features at a given bias.

3. Reducing the applied bias further shifts the band downwards until the "flat band" condition is reached, depicted in Fig. 4b. The flat band situation corresponds to

$$eV_{FB} = E_A + E_{gap} - E_b/2 - \Phi_{tip}$$

Where $e$ is the electron charge, $E_A$ is the electronic affinity of MoS2, $E_{gap}$ is the bandgap width, $E_b$ is the ionization energy of the impurity level, and $\Phi_{tip}$ is the tip work function.

## 4. Summary and conclusions

We investigated native defects on the surface of commercial, high-quality bulk crystals of MoS2 via a combination of scanning probe techniques. STS mappings about the location of an embedded sulfur vacancy provide the first observation of ionization rings on MoS2, induced by the presence of a native defect rather than a substitutional dopant. We follow the spatial and energetic distribution of the defect-induced state in the bulk bandgap of MoS2, and provide a simple electrostatic model to explain the observed dependence on bias and radius as donor ionization caused by TIBB. Our work points out that caution must be taken in interpreting defect-related features in STS of semiconductors, as the position in energy of the in-gap defect-related conductance observed via STS may not correspond to the actual in-gap density of states of the defect state.

## Acknowledgements

I.D.B., L.W., S.C., Y.S.C., M.T.E. and M.S.F. were supported by FLEET Centre of Excellence, ARC grant no. CE170100039. M.T.E. was also supported by ARC DECRA fellowship DE160101157 and M.S.F. by ARC Laureate Fellowship FL120100038. K. X. was supported by Discovery Project DP200101345. The authors acknowledge use of facilities within the Monash X-ray Platform. Part of the device fabrication was performed at the Melbourne Centre for Nanofabrication (MCN) in the Victorian Node of the Australian National Fabrication Facility (ANFF).